\documentclass[12pt]{article}

\usepackage{amsmath,amssymb,graphicx,multicol} 
\usepackage{epsf,color}
\usepackage{pstricks}
\usepackage{cite}

\newcommand{
\beq}{\begin{eqnarray}}
\newcommand{\eeq}{\end{eqnarray}}

\newcommand{\centeron}[2]{{\setbox0=\hbox{#1}\setbox1=\hbox{#2}\ifdim

\wd1>\wd0\kern.5\wd1\kern-.5\wd0\fi
\copy0

\kern-.5\wd0\kern-.5\wd1\copy1\ifdim\wd0>\wd1
                                     \kern.5\wd0\kern-.5\wd1\fi}}
\newcommand{\ltap}{\>\centeron{\raise.35ex\hbox{$<$}}
                             {\lower.65ex\hbox{$\sim$}}\>}
\newcommand{\gtap}{\>\centeron{\raise.35ex\hbox{$>$}}
                             {\lower.65ex\hbox{$\sim$}}\>}

\newcommand\ZZ{\hbox{\zfont Z\kern-.4emZ}}
\font\zfont = cmss10 
\newcommand{\sfrac}[2]{{\textstyle\frac{#1}{#2}}}

\newcommand{\eq}[1]{(\ref{eq:#1})}
\newcommand{\eps}{\epsilon}
\newcommand{\pd}{\partial} 

\begin{document}
\begin{titlepage}
\begin{flushright}
\end{flushright}

\vskip.5cm
\begin{center}
{\huge \bf 
The AdS/CFT/Unparticle}
\vskip.2cm
{\huge \bf  Correspondence}

\vskip.1cm
\end{center}
\vskip0.2cm

\begin{center}
{\bf
{Giacomo Cacciapaglia}${}^{a}$, {Guido Marandella}${}^{b}$,\\
{\rm and}
{John Terning}${}^{b}$}
\end{center}
\vskip 8pt

\begin{center}
{\it
${}^a$Institut de Physique Nucl\'eaire de Lyon, Universit\'e Lyon 1, CNRS/IN2P3,\\
F-69622 Villeurbanne Cedex, France} \\
\vspace*{0.3cm}
{\it ${}^b$ Department of Physics, University of California, Davis, CA  
95616, USA} \\
\vspace*{.5cm}
\vspace*{0.3cm}
{\tt  g.cacciapaglia@ipnl.in2p3.fr, maran@physics.ucdavis.edu, terning@physics.ucdavis.edu}
\end{center}

\vglue 0.3truecm

\begin{abstract}
\vskip 3pt
\noindent
We examine the correspondence between the anti-de Sitter (AdS) description of conformal field theories (CFTs) and the unparticle description of CFTs. We show how unparticle actions are equivalent to holographic boundary actions for fields in AdS, and how massive unparticles provide a new type of infrared cutoff that can be simply implemented in AdS by a soft breaking of conformal symmetry.
We also show that processes involving scalar unparticles with dimensions $d_s<2$ or fermion unparticles with dimensions $d_f<5/2$ are insensitive to ultraviolet cutoff effects.
Finally we show that gauge interactions for unparticles can be described by bulk gauge interactions in  AdS and that they correspond to minimal gauging of the non-local effective action, and we compute the fermion unparticle production cross-section.

\end{abstract}

\end{titlepage}

\newpage


\setcounter{footnote}{0}
\section{Introduction}
\label{sec:intro}
\setcounter{equation}{0}

Recently Georgi \cite{Georgi,Georgi2} proposed a new way of dealing with conformal field theories (CFTs).  The exact propagator of a gauge invariant field is fixed up to a normalization by the scaling dimension of the field.  The corresponding final state phase space looks like that of a non-integer number of particles, and so Georgi dubbed them unparticles. Weakly coupled field theories in  anti-de Sitter (AdS) space provide  another way of approaching large $N$ CFTs by means of the AdS/CFT correspondence~\cite{Maldacena-17}. Since unparticles and AdS theories can in principle both describe the same CFT there must be a way to translate results from one approach into the other.  As is often the case, this is not merely a trivial exercise, since problems that are difficult to solve in one language turn out to be easy to solve in the other.
In this paper we will work through several cases where insights from AdS provide new clarity to the unparticle description and also where insights from the unparticle description lead to new developments in the AdS description.

In \cite{ColoredUnparticles} we proposed using an effective (non-local) action to work with unparticles. This action is completely fixed by requiring that it gives the correct unparticle propagator. Using this effective action it is a relatively straightforward exercise to gauge the global symmetries and to derive the Feynman vertices for gauge interactions~\cite{gauging,Mandelstam,gauging2}. However the resulting inclusive scalar unparticle production cross-section via gauge bosons only makes sense for scaling dimensions below two. If the cross-section is extrapolated to scaling dimensions above two it turns negative. Clearly the effective action cannot be used for scaling dimensions above two, but it would be nice to have a simple physics explanation.  Such an explanation will arise in this paper.

Another issue related to gauge interactions is the question of whether it is theoretically consistent to consider minimal coupling of an unparticle to an ordinary gauge boson, neglecting the possible mixing of the gauge bosons with spin one CFT states with the same gauge quantum numbers.  In the AdS description this mixing typically results in the absence of a normalizable zero mode gauge boson corresponding in 4D to the gauge coupling being driven to zero in the infrared (IR) by the large $N$ CFT.
Finally, unparticles with standard model gauge charges are compatible with low energy experiments in the real world only if their effects are somehow cut off at low energies. Some phenomenological proposals for how to break the unparticle conformal invariance at low energies exist~\cite{Fox,ColoredUnparticles}, however it is not clear if it is possible to do so while preserving a continuum of states above the threshold and without introducing light particle resonances, whose presence would completely spoil the unparticle approximation. Naively one might suspect that introducing an IR cutoff in AdS corresponds to confinement in the CFT and thus produces towers of Kaluza-Klein (KK) modes rather than just cutting off the bottom of a continuum.  In  \cite{ColoredUnparticles}  we suggested that solving the problem of the IR cutoff can also solve the problem of the gauge coupling being driven to zero.  The idea is most simply illustrated by considering a Banks-Zaks model \cite{BanksZaks}.  A large $N$ Yang-Mills theory can be tuned to have a weak IR fixed point by adjusting the number of fermions so that the one-loop $\beta $ function coefficient is very small and the two-loop term has the opposite sign. We can weakly gauge a subgroup of the global symmetry by letting two of the (Weyl) fermions have opposite perturbative $U(1)$ gauge charges. Since the fermions are massless they will drive the $U(1)$ gauge coupling to zero in the IR.  If we give the fermions with the $U(1)$ charges a Dirac mass, then they will no longer contribute to the running below this mass and the $U(1)$ gauge coupling will not run to zero.  The absence of this fermion in the IR will change the value of the Banks-Zaks IR fixed point by an amount of ${\cal O}(1/N)$. Since the massive fermion has a finite anomalous dimension it will look like an unparticle above its mass but the phase space will be cutoff below the mass.  There is no possibility of any confining behavior in this model, so it is an existence proof that we should be able to introduce an IR cutoff for unparticles that does not correspond to confinement.

In this paper we will show how such an IR cutoff can be introduced in AdS and that it reproduces exactly the simple IR cutoff used for unparticles and does not give a spectrum like a confining theory. We will also show how the zero mode gauge boson can be normalizable in AdS without an IR brane.

The outline of the paper is as follows. In Section 2 we extend the unparticle formalism to fermions and show that the results correspond exactly with those of the AdS/CFT correspondence. In Section 3 we show how holographic boundary actions are related by a Legendre transformation to unparticle actions. In Section 4 we describe a new type of IR cutoff for AdS that reproduces the massive unparticle ansatz without KK resonances. The case of scalar unparticle thresholds is discussed completely while some of the details of the fermion unparticle thresholds are left for an appendix.
In Section 5 we show how the bulk gauge action must be modified so that the corresponding 4D $\beta$ function behaves correctly and how this leads to a normalizable zero mode. In Section 6 we discuss  inclusive production cross-sections, what happens when the scalar dimension goes above two, and how the results are extended to unfermions, summarizing the Feynman rules for scalar and fermion unparticles in an appendix. Finally we give our conclusions in Section 7.

\section{Fermion Unparticles and AdS}
\label{sec:unfermions}
\setcounter{equation}{0}

As a warm up, we will first analyze the relation between fermion unparticles and bulk fields in AdS$_5$.
The correspondence between scaling dimensions for  fermions in AdS$_5$ with an ultraviolet (UV) cutoff has been analyzed in Ref.~\cite{ContinoPomarol} and in AdS$_5$ with supersymmetry in \cite{SUSYAdS}, so we will simply sketch the analysis and refer to these references for further details.
We focus on fermions here because the case of scalar fields and unparticles has been already considered in some detail in the literature (see \cite{KlebanovWitten,PerezVictoria} and \cite{Stephanov}); furthermore in the scalar case there is a non trivial transition to dimensions less than two \cite{KlebanovWitten}, which is absent in the fermion case.
We will comment on this issue in the next section because dimensions between one and two are actually the most interesting range for scalar unparticle physics.
Finally, as it was pointed out in \cite{SUSYAdS}, using a supersymmetric formalism with off-shell auxiliary fields, fermions and scalars can be treated in a similar fashion and in particular the subtleties for scalar dimensions below two can be removed.

Let us first define fermion unparticles in analogy to the scalar case~\cite{Georgi}: we take  the propagator of a left-handed fermion operator $\Theta$  with scaling dimension $d_f$ to be given by~\footnote{Note that we could have chosen $\sigma\cdot p + M$ in the numerator of Eq.~\eq{unfermprop}: this would have generated a non-local Dirac (or Majorana) mass that scales like $(- p^2 - i \epsilon)^{d_f-3/2}$ and whose coefficient vanishes in the $d_f \to 3/2$ limit. Our choice is the simplest one, and also what we obtain from a single bulk field in AdS~\cite{ContinoPomarol}.}
\beq
\Delta_f(p,d_f)&\equiv&\int d^4 x \, e^{i px} \langle 0|T {\Theta}(x)  {\Theta}^\dagger(0) |0\rangle\nonumber \\&=&
\frac{A_{d_f-1/2}}{2\pi i}  \int_{0}^\infty (M^2)^{d_f-5/2} \frac{ { \sigma}^\mu p_\mu}{p^2-M^2+i \epsilon}dM^2\nonumber\\
&=& \frac{A_{d_f-1/2}}{2 i \cos d_f \pi}\left({ \sigma}^\mu p_\mu\right) \left(-p^2-i \epsilon \right)^{d_f-5/2} + \dots~\,.
\label{eq:unfermprop}
\eeq
This is, of course, is the form one finds for a massless fermion propagator after resumming collinear emissions of 
a massless gauge boson \cite{Brown,Neubert}.
We can fix the normalization~\cite{luozhu}
\beq
A_{d_f-1/2}=\frac{16\pi^{5/2}}{(2\pi)^{2d_f-1}}
\,\frac{\Gamma(d_f)}{\Gamma(d_f-3/2)\,\Gamma(2d_f-1)}
\label{Ad}
\eeq
so that in the limit $d_f\rightarrow 3/2$ we find 
\beq
\Delta_f(p,3/2)
&=& \frac{i { \sigma}^\mu p_\mu}{p^2+i \epsilon}~,
\eeq
and we recover the usual particle propagator for a left-handed fermion. Note that $d_f-1/2=$ $d_f$~$-$~spin is what is conventionally known as the ``twist" of the operator. Let us finally recall that for a right-handed unparticle we simply replace ${\sigma}^\mu p_\mu$ with ${\bar \sigma}^\mu p_\mu$. 
Taking the discontinuity across the cut from Eq. (\ref{eq:unfermprop}) we find the phase space factor
\beq
d\Phi_f(p,d_f)=A_{d_f-1/2}\,\theta\left(p^0 \right)
\,\theta\left(p^2 \right) ({ \sigma}^\mu p_\mu) \,(p^2)^{d_f-5/2}~
\eeq
which, in the limit $d_f\rightarrow 3/2$, becomes the usual phase space factor for a massless fermion:
\beq
d\Phi_f(p,3/2)= 2 \pi \; \theta\left(p^0 \right) \left({ \sigma}^\mu p_\mu \right)\, \delta(p^2)~.
\eeq

Note however that the propagator in Eq.~(\ref{eq:unfermprop}) is not well defined for $d_f>5/2$: in fact, for large scaling dimensions, the integral is divergent and the propagator becomes UV sensitive.
Therefore (just as with scalars with $d_s>2$) other terms, that depend on the UV completion of the theory, must be added to the propagator (and are represented by the dots in Eq.~(\ref{eq:unfermprop})).
Another way to formulate the problem is that for $d_f>5/2$, Georgi's propagator is dominated by large momenta.
On the other hand, the phase space derived from such propagator, is well behaved for all dimensions ($d_f>3/2$).
We will comment more on this point by the end of this section.

Let us now compare the fermion unparticle analysis to fermions in AdS$_5$.
We will use the conformally flat metric 
\beq
ds^2 
= {{R^2}\over{z^2}}\left(\eta_{\mu \nu}dx^\mu dx^\mu-dz^2 \right)~,
\label{conformallyflat}
\eeq
with a brane at $z=\epsilon$, which represents a UV cutoff $\Lambda_{UV} \sim 1/\epsilon$, and the space extending to infinity~\cite{RandallSundrum2}. 
We write the 5D Dirac fermion ${\bf \Psi}$ in terms of two Weyl spinors $\chi$ and  $\bar \psi$ 
\beq 
{\bf \Psi}=\left( \begin{array}{c} \chi \\ \bar \psi \end{array} \right)\,.
\eeq
The 5D Lagrangian is
\begin{equation} 
\mathcal{L}_{5D}  = \int_\epsilon^\infty d z 
\left(\frac{R}{z}\right)^4 
\left( 
- i \bar{{\bf \chi}}  {\bar \sigma}^\mu \partial_\mu {\bf \chi} 
- i \psi  \sigma^{\mu} \partial_\mu \bar{\psi} 
+ \sfrac{1}{2} ( \psi \overleftrightarrow{\partial_z} {\bf \chi} 
-  \bar{{\bf \chi}}  \overleftrightarrow{\pd_z} \bar{\psi} )
+ \frac{c}{z} \left( \psi {\bf \chi} + \bar{{\bf \chi}} \bar{\psi} \right) 
\right)\,, 
\end{equation} 
where $\overleftrightarrow{\partial_z}  = \overrightarrow{\partial_z}-\overleftarrow{\partial_z}$
with the convention that the differential operators act only on the spinors and not on the metric factors. 
The bulk equations of motion (EOMs) can be solved by
\begin{equation} 
{\bf \chi} 
= 
g(p, z)\, \chi_4
\ \ \mathrm{and} \ \ 
\bar{\psi} 
=
f (p,z)\, \bar{\psi}_4, 
\end{equation} 
where the 4D plane-wave spinors $\chi_4$ and $\bar{\psi}_4$ satisfy the usual 4D Dirac equation with mass $p=\sqrt{p^2}$:
\begin{equation} 
\label{eq:Dirac}
-i \bar{\sigma}^\mu \partial_\mu \chi_4 + p\, \bar{\psi}_4= 0
\ \ \mathrm{and} \ \ 
-i \sigma^\mu \partial_\mu \bar{\psi}_4 + p\,\chi_4 = 0. 
\end{equation} 
The solutions are linear combinations of Bessel functions
\begin{eqnarray} 
      \label{eq:Bessel1}
& \displaystyle
g(p, z) 
=
A\, z^\frac{5}{2} 
\left( 
c_\alpha  J_{c+\frac{1}{2}}(p z) + s_\alpha J_{-c-\frac{1}{2}}(p z) 
\right)\,,
\\ 
      \label{eq:Bessel2}
& \displaystyle
f(p, z) 
= 
A\, z^\frac{5}{2} 
\left( c_\alpha J_{c-\frac{1}{2}}(p z) - s_\alpha J_{-c+\frac{1}{2}}(p z)
\right)\,;
\end{eqnarray} 
where $c_\alpha^2 + s_\alpha^2 = 1$, and we are assuming, for simplicity, that $c\pm 1/2$ is not an integer.
The ratio $s_\alpha/c_\alpha$ is determined by the boundary conditions (BCs) in the IR, $z \to \infty$, and we will leave it undetermined for the moment. 
In fact, the properties of the CFT operator do not depend on this choice.
The normalization $A$ is determined by the BCs on the UV brane: this depends on which source we couple the CFT operator to, and also determines the chirality of the CFT operator.

For instance, a left-handed source coupled to a right-handed CFT operator corresponds to fixing the value of the left-handed bulk field on the UV brane:
\beq
\chi(p, \eps) = \chi_0 = g_0(p)\, \chi_4\,,
\eeq 
where $\chi_0$ is the source field.
This BC fixes the normalization factor $A$:
\beq
A = \frac{g_0\,\epsilon^{-5/2}}{c_\alpha  J_{c+\frac{1}{2}}(p \epsilon) + s_\alpha J_{-c-\frac{1}{2}}(p \epsilon)}\,.
\eeq
After using the bulk EOMs, the bulk action reduces to a pure boundary term on the UV brane:
\beq
\mathcal{L} & = & \frac{1}{2} \left( \frac{R}{\epsilon} \right)^4 \, \chi(\epsilon) \psi(\epsilon) + h.c. = \frac{1}{2} \left( \frac{R}{\epsilon} \right)^4 \, f(p,\epsilon) g(p,\epsilon)\, \chi_4 \psi_4 + h.c. \nonumber\\
& = & \left( \frac{R}{\epsilon} \right)^4 \frac{c_\alpha  J_{c-\frac{1}{2}}(p \epsilon) - s_\alpha J_{-c+\frac{1}{2}}(p \epsilon)}{c_\alpha  J_{c+\frac{1}{2}}(p \epsilon) + s_\alpha J_{-c-\frac{1}{2}}(p \epsilon)}\, \frac{\bar \chi_0 {\bar \sigma}^\mu p_\mu \chi_0}{p}\,,\,
\eeq
where we have used Eqs.~(\ref{eq:Dirac}).
Since $\chi_0$ is  the source for the CFT operator, we can identify the 2 point correlator of the CFT operator with the kinetic term of $\chi_0$:
\beq
\langle \Theta_R \Theta_R \rangle = \Delta_R (p, c) \sim \left( \frac{R}{\epsilon} \right)^4 \frac{c_\alpha  J_{c-\frac{1}{2}}(p \epsilon) - s_\alpha J_{-c+\frac{1}{2}}(p \epsilon)}{c_\alpha  J_{c+\frac{1}{2}}(p \epsilon) + s_\alpha J_{-c-\frac{1}{2}}(p \epsilon)}\, \frac{\bar \sigma^\mu p_\mu}{p}\,.
\eeq

In order to study the properties of the CFT operator, we need to recover exact conformal invariance by removing the UV cutoff.
We can now expand the Bessel functions for large UV cutoff, $p \epsilon \ll 1$:
\beq \label{eq:besselexpansion}
J_\nu (p \epsilon)= (p \epsilon)^\nu \sum_{n=0}^{\infty} (p \epsilon)^{2 n} \frac{(-1)^n}{2^{2n+\nu} \Gamma(n+1) \Gamma(n+\nu+1)} \equiv (p \epsilon)^\nu \; \Sigma_\nu (p \epsilon)~,
\eeq
where $\Sigma_\nu$ only contains positive and integer powers of $p^2$.

In the case $c > -1/2$, we find that the propagator splits into a local and a non-local term
\beq
\Delta_R  = \Delta_{R,(local)} + \Delta_{R,(non-local)}\,. \label{eq:locnonloc}
\eeq
After rescaling $\chi_0$ by $R^{-2} \epsilon^{2-c}$, we find~\footnote{An expansion of ratios of $\Sigma$'s for small arguments is implicitly assumed in this formulas (and in the following). In this sense $\Delta_{(local)}$ contains (infinite) local terms.}:
\beq
\Delta_{R,(local)} &=&- \epsilon^{1-2c}\, (\bar \sigma \cdot p)\, \frac{\Sigma_{-c+1/2} (p\epsilon)}{\Sigma_{-c-1/2} (p\epsilon)}\,, \label{eq:Dlocal} \\
\Delta_{R,(non-local)} & = & \frac{c_\alpha}{s_\alpha} \frac{\bar \sigma \cdot p}{p^{1-2c}} \left[ \left( \frac{\Sigma_{c-1/2} (p\epsilon)}{\Sigma_{-c-1/2} (p\epsilon)} + (p\epsilon)^2 \frac{\Sigma_{c+1/2} (p\epsilon)\, \Sigma_{-c+1/2} (p\epsilon)}{\Sigma^2_{-c-1/2} (p\epsilon)}\right)   \right.  \label{eq:Dnonlocal}\\
& &\quad\quad\quad\quad\quad \left. + \phantom{\frac{\Sigma_1}{\Sigma_2}} \mathcal{O} (\epsilon^{2c+1}) \right]\,. \nonumber
\eeq

From Eq.~(\ref{eq:Dnonlocal}) we can deduce that the dimension of the CFT operator is
\beq
d_{fR} = 2+c \qquad \mbox{for} \quad c>-1/2\,,
\eeq
in agreement with Refs~\cite{ContinoPomarol,SUSYAdS}.
In the limit $\epsilon \to 0$, the non-local part of the propagator is equal to the unfermion propagator (\ref{eq:unfermprop}), up to a normalization which depends on the boundary conditions in the IR.

Now consider the local terms. For $-1/2 < c < 1/2$, the local terms are subleading, i.e. they will vanish in the limit $\epsilon \to 0$.
However, for $c>1/2$, when the dimension of the CFT operator is $d_{fR} > 5/2$, some of the local terms are enhanced by negative powers of $\epsilon$ and they diverge in the limit $\epsilon \to 0$.
Such terms are equivalent to the UV sensitivity of the unfermion propagator (\ref{eq:unfermprop}) for $d_f>5/2$: indeed, one has to add counterterms on the UV brane to cancel them\footnote{Note that for a given $d_f$, only a finite number of counterterms is necessary.}.
Those counterterms only affect the propagator of the unparticle, but not the phase space.
In fact, when the unparticle propagator mediates an interaction between ordinary particles, those local terms will generate effective higher dimension operators.
Note also that the coefficient of the local term is independent of the IR physics and it is completely determined in terms of the AdS$_5$ geometry.

In the region $c<-1/2$, the propagator can be expanded as
\beq
\Delta_R (p) \sim \frac{\bar \sigma \cdot p}{p^2}\, \frac{\Sigma_{c-1/2} (p \epsilon)}{\Sigma_{c+1/2} (p \epsilon)} + \mathcal{O} (\epsilon^{-1-2c})\,. \label{eq:Dpole} \eeq
In the conformal limit, $\epsilon \to 0$, we recover the propagator of a massless particle: in this region, therefore, the theory describes a free massless fermion \cite{SUSYAdS}.
Note that this interpretation differs from the one in Ref.~\cite{ContinoPomarol} where the authors are considering a theory with a fixed UV cutoff rather that taking the limit $\epsilon \rightarrow 0$ to remove the cutoff and recover  pure CFT behavior.

We can repeat the same analysis for a left-handed CFT operator by changing the UV BCs: this corresponds to a different CFT. 
The physics is again different from Ref.~\cite{ContinoPomarol} where the authors interpreted the two BCs  to correspond to equivalent CFT theories up to UV localized degrees of freedom:  when the UV brane is removed, this interpretation is not possible.
The relevant BC now is
\beq
\psi(p, \eps) = \psi_0 = f_0(p)\, \psi_4\,,
\eeq 
where $\psi_0$ is the right-handed source field.
In this case, the CFT propagator is
\beq
\langle \Theta_L \Theta_L \rangle = \Delta_L (p, c) \sim \left( \frac{R}{\epsilon} \right)^4 \frac{c_\alpha  J_{c+\frac{1}{2}}(p \epsilon) + s_\alpha J_{-c-\frac{1}{2}}(p \epsilon)}{c_\alpha  J_{c-\frac{1}{2}}(p \epsilon) - s_\alpha J_{-c+\frac{1}{2}}(p \epsilon)}\, \frac{\sigma_\mu p^\mu}{p}\,.
\eeq
For $c<1/2$ we find, after rescaling $\psi_0$:
\beq
\Delta_{L,(local)} &=& \epsilon^{1+2c}\, (\sigma \cdot p)\, \frac{\Sigma_{c+1/2} (p\epsilon)}{\Sigma_{c-1/2} (p\epsilon)}\,, \label{eq:DlocalLH} \\
\Delta_{L,(non-local)} & = & \frac{s_\alpha}{c_\alpha} \frac{\bar \sigma \cdot p}{p^{1+2c}} \left[ \left( \frac{\Sigma_{-c-1/2} (p\epsilon)}{\Sigma_{c-1/2} (p\epsilon)} + (p\epsilon)^2 \frac{\Sigma_{c+1/2} (p\epsilon)\, \Sigma_{-c+1/2} (p\epsilon)}{\Sigma^2_{c-1/2} (p\epsilon)}\right)   \right.  \label{eq:DnonlocalLH}\\
& &\quad\quad\quad\quad\quad \left. + \phantom{\frac{\Sigma_1}{\Sigma_2}} \mathcal{O} (\epsilon^{1-2c}) \right]~. \nonumber
\eeq
Therefore, the dimension of the left-handed operator is
\beq
d_{fL} = 2-c\,, \qquad \mbox{for} \quad c<1/2\,,
\eeq
and the local terms are relevant when $c<-1/2$ (and $d_{fL} > 5/2$).
Again, for $c>1/2$ we get a free massless fermion.
Note finally the relation:
\beq
\Delta_R (p, -c) = - \Delta_L (p, c)\,.
\eeq

\section{Holographic Boundary Actions and Unparticle Actions}
\label{sec:actions}
\setcounter{equation}{0}

In the previous section we showed how to relate boundary actions to unparticle propagators.
Boundary actions can also be used as effective unparticle actions in some cases. Consider for instance a real bulk scalar in AdS$_5$:
\beq
S_{bulk}= {{1}\over{2}}\int d^4 x \, dz \left( \frac{R}{z}\right)^3 \left(\partial_M \phi
\partial^M \phi + \frac{m^2 R^2}{z^2} \phi^2 \right)~.
\label{AdSscalaraction}
\eeq
We know from the AdS/CFT correspondence that this field corresponds to a CFT operator  ${\mathcal O}$  with scaling dimension
\beq
d_s = 2 \pm \nu=2\pm \sqrt{4 + m^2 R^2}~.
\label{scalingdim}
\eeq
For $z\sim \eps$ the bulk solution of the equation of motion takes the form:
\beq
\phi(p,z)\approx \phi_0(p) (pz)^{2+\nu}+ \beta(p) (pz)^{2-\nu}~.
\eeq
In the AdS/CFT correspondence $\phi_0(p)$ is associated with the source of the operator 
${\mathcal O}$ while $\beta(p)$ is associated with the VEV of ${\mathcal O}$.
Plugging the solution into the bulk action, and integrating by parts one finds a holographic
boundary action which depends on $\phi_0$:
\beq
S_{holo}= {{1}\over{2}}\int \frac{d^4 p}{(2\pi)^4} \phi_0(p) p^{2 \nu} \phi_0(p)~.
\label{holographicBA}
\eeq
For $d_s>2$  the two-point function of the corresponding operator ${\mathcal O}$ is
\beq
\langle {\mathcal O}(p^\prime) {\mathcal O}(p) \rangle 
&\propto&  {{\delta^2 S_{holo}}\over{\delta \phi_0(p^\prime) \, \delta \phi_0(p) }}
\propto \frac{\delta^{(4)}(p+p^\prime)}{ (2 \pi)^4} \, (p^2)^{d_s-2} ~.
\eeq
For $d_s<2$ one must perform a Legendre transformation \cite{KlebanovWitten} which interchanges the source and the field:
\beq
S^\prime_{holo}[\beta]=S_{holo}[\phi_0]+{{1}\over{2}}\int \frac{d^4 p}{(2\pi)^4} \phi(p) \beta(p)
\eeq
and the two-point function of the corresponding operator ${\mathcal O}$ is 
\beq
\langle {\mathcal O}(p^\prime) {\mathcal O}(p) \rangle &\propto&  {{\delta^2 S^\prime_{holo}}\over{\delta \beta(p^\prime) \, \delta \beta(p) }}
\propto \frac{\delta^{(4)}(p+p^\prime)}{ (2 \pi)^4} \, (p^2)^{d_s-2} ~.
\eeq
In both cases we find the correct scaling for the two-point function of an operator of dimension $d_s$ in a 4D CFT. The Legendre transformation interchanges the roles of $\phi_0$ and $\beta$, $\beta$ is now the source and $\phi_0$ corresponds to the VEV.
Thus for $d_s<2$ we have that the propagator of ${\mathcal O}$ and the propagator $\Delta_s(p)$ of the boundary field $\phi_0$ are proportional:
\beq
\langle {\mathcal O}(p^\prime) {\mathcal O}(p) \rangle &\propto&  \frac{\delta^{(4)}(p+p^\prime)}{ (2 \pi)^4} \, (p^2)^{d_s-2}  =\Delta_s(p)~,
\eeq
whereas they are inversely related for $d_s>2$.

As with fermions, there is a direct correspondence between the AdS/CFT description given above and the unparticle/CFT description of Georgi \cite{Georgi}.  For a scaling dimension $d_s$ where $1\le d_s<2$ we have
\beq
\Delta_s(p,d_s)&\equiv& \int d^4 x \, e^{i px} \langle 0|T {\cal  O}(x)  {\cal O}^\dagger(0) |0\rangle 
\nonumber\\
&=&
\frac{A_{d_s}}{2\pi}  \int_{0}^\infty (M^2)^{d_s-2} \frac{i }{p^2-M^2+i \epsilon}dM^2 \nonumber \\
&=&
\frac{A_{d_s}}{2 \sin d_s \pi} \frac{i}{\left( -p^2-i \epsilon \right)^{2-d_s} }~.
\label{eq:unparticle}
\eeq

Thus, up to a normalization, we can identify the the holographic boundary action (\ref{holographicBA}) with an effective action for a scalar unparticle for $d_s<2$.
There is a similar effective action for fermions with scaling dimensions $d_f<5/2$.  Consider a left-handed CFT Weyl fermion operator $\Theta$ with dimension $d_f$ which couples to a right-handed fermion source, ${\bar \psi}_0$.  In the AdS/CFT correspondence
we consider a bulk Dirac fermion made out of two Weyl fermions: $\chi$ which is left-handed and ${\bar \psi}$ which is right-handed with a bulk mass $c$.  
As described in section \ref{sec:unfermions}, after using the equations of motion and imposing the   boundary condition 
\beq
\chi(p,\eps)=0 \ \ \mathrm{and} \ \  \psi(p,\eps)=\psi_0(p)~,
\eeq
we find a boundary holographic action
\beq
S_{holo,f}= {{1}\over{2}}\int \frac{d^4 p}{(2\pi)^4} \psi_0\left[ \,i \sigma^{\mu}  p_\mu  (p^2)^{d_f-5/2} \right]\bar{\psi_0}~.
\eeq
This of course implies that $\psi_0$ has dimension $4-d_f$.
The two-point function of the CFT fermion is given by
\beq
\langle \Theta(p^\prime) \Theta(p) \rangle &\propto& \frac{\delta^2 S_{holo,f}}{\delta \psi_0(p^\prime) \, \delta {\bar \psi}_0(p) }
\propto \frac{\delta^{(4)}(p+p^\prime)}{ (2 \pi)^4} \,  i \sigma^{\mu}  p_\mu (p^2)^{d_f-5/2} ~.
\eeq
Now we can rewrite the boundary holographic action in terms of a the left-handed fermion $\chi_0$ with dimension $d_f$ by performing a Legendre transformation in order to interchange the source and the field:
\beq
S^\prime_{holo,f}= S_{holo,f}+\int \frac{d^4 p}{(2\pi)^4} \left[ \chi_0 \psi_0 +{\bar  \chi}_0 {\bar \psi}_0 
\right]~.
\eeq
Solving for $\psi_0$ we find
\beq
\sigma^\mu p_\mu (p^2)^{d_f-5/2}  \,\bar{\psi}_0 + \chi_0 &=& 0~, \\
\psi_0 \, \sigma^\mu p_\mu (p^2)^{d_f-5/2}   + {\bar \chi}_0 &=& 0~. 
\eeq
This gives the action in terms of a left-handed fermion

\beq
S^\prime_{holo,f}= {{1}\over{2}}\int \frac{d^4 p}{(2\pi)^4}  \bar{\chi_0}\left[  i \bar{\sigma}^\mu p_\mu   (p^2)^{3/2-d_f} \right] \chi_0~.\label{eq:holof}
\eeq
As in the scalar case, the two point function of the left-handed CFT fermion is proportional to the propagator, $\Delta_f(p)$, of the field $\chi_0$ 
\beq
\langle \Theta(p^\prime) \Theta(p) \rangle &\propto&  \frac{\delta^{(4)}(p+p^\prime)}{ (2 \pi)^4} \, \frac{ i}{ \bar{\sigma}^{\mu}  p_\mu (p^2)^{3/2-d_f} }=\Delta_f(p)~.
\eeq
So we see that the boundary holographic action 
$S^\prime_{holo,f}$ approaches the free fermion action as $d_f \rightarrow 3/2$ and can be used as an effective action for the left-handed CFT fermion when $d_f<5/2$.

\section{The IR Cutoff: Soft Breaking of Conformal Invariance}
\label{sec:IRcutoff}
\setcounter{equation}{0}

In order to make unparticles with Standard Model (SM) gauge interactions phenomenologically realistic, it is important to break the conformal symmetry so that the unparticle does not propagate at low invariant mass squared.
Another reason is that the SM has a natural source of conformal symmetry breaking in the Higgs VEV~\cite{Fox,Quiros,Lee}.
Some phenomenological models of conformal breaking that do preserve the continuum at high energies have already been proposed in the literature for scalar unparticles~\cite{ColoredUnparticles,Fox}: the idea is to modify the propagator so that the continuum spectrum is cut off below an IR scale $\mu$
\beq
\Delta_s(p,\mu,d_s)&=&
i \frac{A_{d_s}}{2\pi}  \int_{\mu^2}^\infty (M^2-\mu^2)^{d_s-2} \frac{1}{p^2-M^2+i \epsilon}dM^2\nonumber\\
&=& i\frac{A_{d_s}}{2 \sin d_s \pi} \left( \mu^2-p^2-i \epsilon \right)^{d_s-2} + \dots~.
\eeq
This model can be easily extended to fermion unparticles: the modified propagator is
\beq
\Delta_f(p,\mu,d_f)&=&
\frac{A_{d_f-1/2}}{2\pi i}  \int_{\mu^2}^\infty (M^2-\mu^2)^{d_f-5/2} \frac{ \not p +\mu }{p^2-M^2+i \epsilon}dM^2\nonumber\\
&=& \frac{A_{d_f-1/2}}{2 i \cos d_f \pi}\left(\not p +\mu\right) \left( \mu^2-p^2-i \epsilon \right)^{d_f-5/2} + \dots~,
\label{eq:IRprop}
\eeq
where $\not p =\gamma^\alpha p_\alpha$.  The fermion propagator has the form found for a massive fermion after resumming collinear emissions of 
a massless gauge boson \cite{Brown}.

In AdS$_5$, the simplest way to introduce an IR cutoff is to cut off the space at large $z$ with an IR brane~\cite{RS}.
However, this cutoff would turn the CFT into a confining theory and the spectrum would be a tower of massive states whose spacing in mass is given by the IR cutoff itself.
Can we implement a simple IR cutoff in AdS that preserves the continuum above the scale $\mu$?
This would be easily achieved if the propagator were a function of $\sqrt{p^2 - \mu^2}$ instead of $\sqrt{p^2}$: this happens in flat space, given a bulk mass $\mu$.
In AdS space, in order to generate a shift of the  $p^2$ dependence, we need to add a mass term with a profile along the extra dimension to compensate for the warp factors, for instance generated by the VEV of a bulk scalar.
In the case of a scalar unparticle this is simple to do.
We can add a bulk interaction with a new field $H$: $H\phi \phi$. If $H$ has scaling dimension 2 then its profile will scale like $z^2$ with an  appropriate choice of boundary conditions for $H$.
With the background field solution $H=\mu^2 z^2$ the equation of motion for $\phi$ (with 4-momentum $p$) is:
\beq
z^5 \partial_z \left( {{1}\over{z^3}} \partial_z \phi \right) -z^2(p^2-\mu^2) \phi- m^2 R^2 \phi=0~.
\label{Chap17:scalareqm}
\eeq
With the boundary condition $\phi(p,\eps)=\phi_0 (p)$ the solution is
\beq \label{eq:solfield2}
\phi(p, z)=\phi_0(p) \left(\frac{z}{\eps}\right)^2 \frac{a J_\nu(E z) + b J_{- \nu}(E z) }{a J_\nu(E \eps) + b J_{- \nu}(E \eps)}~,
\eeq
where $E=\sqrt{p^2-\mu^2}$.  It is now clear that if we had chosen a different power for the IR growth of the $H$ VEV we could implement a different type of IR cutoff, however other choices would lead to much more complicated bulk equations of motion, which would not, in general, have such simple solutions. Adding a mass term to a CFT is a soft breaking of conformal symmetry that does not change the UV behavior and this calculation seems to show that in the AdS description this soft breaking corresponds to a spontaneous breaking of the symmetry.
With the simple choice we have made the effective  mass term scales with the same power of $z$ as the kinetic term.  Thus in the absence of a  VEV
for $H$, the term $H\phi^2$ scales like the kinetic term for $\phi$ and preserves the $SO(4,2)$ isometry that corresponds to the conformal symmetry of the 4D theory.

For  unfermions things are more complicated.
A mass term for a single bulk fermion is still not enough: we need to have two bulk fields, $\Psi_L$ with bulk mass $c_L$ and $\Psi_R$ with bulk mass $c_R$, and add a cross term like
\beq
\delta \mathcal{L}_{bulk} = \int_\epsilon^\infty dz \left( \frac{R}{z} \right)^5 \lambda H  \left( \chi_L \psi_R + \chi_R \psi_L + h.c. \right)\,.
\eeq
After the scalar $H$ develops a VEV, $\lambda R \langle H \rangle = v(z)$, this term will generate a z-dependent mass term mixing the two bulk fermions.
If $v (z) = \mu z$, the bulk wave functions depend on $E = \sqrt{p^2 - \mu^2}$, provided that $c_L = -c_R \equiv c$.
The interpretation is clear: the two bulk fields are two CFT operators $\Theta_L$ ($\Psi_L$) and $\Theta_R$ ($\Psi_R$) with opposite chirality but the same anomalous dimension $2-c$ thanks to the relation between $c_L$ and $c_R$.
$H$, on the other hand, is a scalar field with dimension 1 whose VEV breaks the conformal invariance and generates a mass term $\mu \Theta_L \Theta_R$.
After imposing the two boundary conditions
\beq
\psi_L (\epsilon) = \psi_0 = f_{0} \psi_4\,, \quad \mbox{and} \quad \chi_R (\epsilon) = \chi_0 = g_{0} \chi_4\,,
\eeq
we can calculate the boundary action (see Appendix~\ref{app:calculation} for details)
\beq \label{eq:LbraneIR}
\mathcal{L}_{boundary} & = & \left( \frac{R}{\epsilon} \right)^4 \left\{  \frac{c_{\alpha L} J_{c+1/2} (E \epsilon) - s_{\alpha L} J_{-c-1/2} (E \epsilon)}{c_{\alpha L} J_{c-1/2} (E \epsilon) + s_{\alpha L} J_{-c+1/2} (E \epsilon)}\; \frac{\psi_0\, \sigma \cdot p\, \bar \psi_0}{E} \right. \nonumber \\
& & \quad\quad\quad\quad -  \frac{s_{\alpha R} J_{c+1/2} (E \epsilon) - c_{\alpha R} J_{-c-1/2} (E \epsilon)}{s_{\alpha R} J_{c-1/2} (E \epsilon) + c_{\alpha R} J_{-c+1/2} (E \epsilon)}\; \frac{\bar \chi_0\, \bar \sigma \cdot p\, \chi_0}{E}  \\
& & \quad\quad\quad\quad + \left. \frac{1}{2} \frac{\mu}{E} M( E\epsilon) \left( \psi_0 \chi_0 + h.c. \right) \right\}\,,\nonumber
\eeq
where
\beq
M(E\epsilon)=
\frac{c_{\alpha L} J_{c+1/2} (E \epsilon) - s_{\alpha L} J_{-c-1/2} (E \epsilon)}{c_{\alpha L} J_{c-1/2} (E \epsilon) + s_{\alpha L} J_{-c+1/2} (E \epsilon)} - 
\frac{s_{\alpha R} J_{c+1/2} (E \epsilon) - c_{\alpha R} J_{-c-1/2} (E \epsilon)}{s_{\alpha R} J_{c-1/2} (E \epsilon) + c_{\alpha R} J_{-c+1/2} (E \epsilon)} ~.
\eeq
For $c<1/2$, we can expand the Bessel functions for small argument, $E \epsilon \ll 1$, as in section \ref{sec:unfermions}.
The local terms, dominant for $c<-1/2$ ($d_f>5/2$) are:
\beq
\mathcal{L}_{local} = \epsilon^{1+2c} \frac{\Sigma_{c+1/2} (E \epsilon)}{\Sigma_{c-1/2} (E \epsilon)}\; \left( \psi_0\, \sigma \cdot p\, \bar \psi_0 -  \bar \chi_0\, \bar \sigma \cdot p\, \chi_0 \right)\,,
\eeq
where we have rescaled $\psi_0$ and $\chi_0$ by a factor $R^{-2} \epsilon^{2+c}$.
Note that in the local terms  the mass $\mu$ cancels out: the conformal breaking  is generated in the IR, and it does not affect the UV physics.
The non-local part is proportional to
\beq
E^{2 c -1} \left( \frac{s_{\alpha L}}{c_{\alpha L}}\, (\psi_0\, \sigma \cdot p\, \bar \psi_0) - \frac{c_{\alpha R}}{s_{\alpha R}}\, (\bar \chi_0\, \bar \sigma \cdot p\, \chi_0) + \frac{1}{2} \left( \frac{s_{\alpha L}}{c_{\alpha L}} -  \frac{c_{\alpha R}}{s_{\alpha R}}  \right)\, \mu \, \left(  \psi_0 \chi_0 + h.c. \right)  \right),
\eeq
and, if we normalize the two kinetic terms properly, i.e. $s_{\alpha L}/c_{\alpha L} = - c_{\alpha R}/s_{\alpha R}$, we can rewrite it as
\beq
\mathcal{L}_{non-local} \sim E^{2 c-1}\, \bar \Psi_0 (\not p + \mu) \Psi_0\,, \quad \mbox{where} \quad \Psi_0 = \left( \begin{array}{c} \chi_0 \\
\bar \psi_0 \end{array} \right)\,.
\eeq
The corresponding propagator coincides with Eq.~(\ref{eq:IRprop}) with $d_f=2-c$.

\subsection{Stability of the AdS geometry in the IR}

A bulk scalar field with a VEV that grows with $z$ can provide an effective IR cutoff for unparticles in AdS$_5$, preserving the continuum spectrum, typical of an unparticle, for momenta larger that the IR cutoff.
This behavior has to be compared to the traditional Randall-Sundrum approach~\cite{RS}, where a brane is used to cut-off the space at large $z$: in the latter case a mass gap is introduced and the spectrum reduces to a series of KK resonances spaced by a mass scale $\propto 1/z_{IR}$.
As we have seen above, a simple model of an unparticle IR cutoff is achieved with a linearly growing VEV for fermions and a quadratic one for scalars.
However, the growing VEV will, at large $z$, destabilize the AdS geometry and its back-reaction will eventually cut off the space and possibly reintroduce a mass gap.
Our description of the model is therefore safe only if such effects appear at a scale much lower than the IR cutoff $\mu$, i.e. $z_{IR} \ll 1/\mu$.

In order to address this issue, we will cutoff the space at $z_{IR}$ by hand and push this brane to the largest possible $z$.
In 5D the strength of gravitational interactions are given by the inverse of the third power of the 5D Planck scale, $M_*^{-3}$.  The curvature of the AdS$_5$ bulk is  \cite{RS}
\beq
\Lambda=-24 \frac{M_*^3}{R^2}
\eeq
and classical solutions for the metric should be a good approximation as long as 
\beq
R \gg \frac{1}{M_*}~.
\eeq
With a scalar VEV for $H$ growing in the IR like a power, i.e.  $\mu z^2/R^3$ and its size controlled by a potential on the IR brane at  $z_{IR}$, there are contributions from brane terms  $\sim \mu^2 H^2$ to the stress-energy tensor that grow in the IR like
\beq
T^{MN}=\frac{2}{\sqrt{-g}}\frac{\delta S}{\delta g_{MN}}\sim \eta^{MN}\, \left(\frac{z_{IR}}{R}\right)^{2}\,\frac{\mu^4}{R^6} z_{IR}^5~.
\eeq
For stability we need this contribution to be small compared to the contribution from the bulk curvature:
\beq
g^{MN} \Lambda = \eta^{MN}  \left(\frac{z}{R}\right)^{2}\,\Lambda~,
\eeq
thus we require 
\beq
\frac{\mu^4}{R^6} z_{IR}^5& \ll&  \frac{24 \,M_*^3}{R^2} ~,\nonumber \\
z_{IR}& \ll& \frac{1}{\mu}\, \left( 24\,R^4  \mu M_*^3 \right)^{1/5}~.
\label{zIRbound}
\eeq
As long as $\mu$ is sufficiently small compared to $1/R$ 
then Eq.~(\ref{zIRbound}) is consistent with 
\beq
\mu \gg \frac{1}{z_{IR}}~.
\eeq
In this case the spacing between KK modes introduced by the IR brane is tiny compared to the mass gap $\mu$.
For example with 
$M_*=10^{19}$ GeV, $R^{-1}= 10^9$ GeV and $\mu =100$ GeV we have 
\beq
\frac{1}{z_{IR}}\gg 10^{-23}\, {\rm GeV}~,
\eeq
so that the minimum spacing between the modes can be $10^{25}$ times smaller than the threshold $\mu$.  This would be a very good approximation to a continuum.

\section{Gauge Interactions}
\setcounter{equation}{0}

A method for  adding  gauge interactions for unparticle fields was proposed in Ref.~\cite{ColoredUnparticles} (see also Refs.~\cite{GaugeUnparticles} for a discussion on gauging unparticles).
In this section we will study how to reproduce the same results in the context of an infinite AdS space: the natural choice would be to gauge a global symmetry by adding a bulk gauge field.
It is well known that such a gauge field corresponds to a vector operator  of dimension 3 (i.e a conserved current)  in the CFT theory.
However, due to the conformal invariance, the associated  external gauge coupling runs to zero in the IR due to a logarithmic dependence on the momentum in the propagator. 
Moreover in the 5D description the zero mode gauge boson is non-normalizable as its normalization is proportional to the volume of the space: in order to have a realistic description of ordinary gauge interactions we need a way to effectively cut off the space without introducing an IR brane.

One possibility is to add a dilaton factor $\Phi(z)$ in front of the kinetic term:
\beq
-\frac{1}{4 g_5^2} \int_\epsilon^\infty d^4x \,dz  \left(\frac{R}{z}\right) \Phi(z) F^{aMN}F^a_{MN}\,. \label{eq:dilaton}
\eeq
The spectrum still contains a flat zero mode, however the relation between 4D and 5D couplings changes.
For a dilaton background that falls sufficiently fast for large $z$, the 4D gauge coupling can be finite.  
As a particular example consider
\beq
\Phi(z)=e^{-m z}~.
\eeq
In this case we have
\beq
\frac{1}{g_4^2}=\frac{R}{g_5^2}\int_\epsilon^\infty \frac{e^{-m z}}{z}=\frac{R}{g_5^2}[{\rm Ei}(-m \infty)-{\rm Ei}(-m \epsilon)]\approx \frac{R}{g_5^2}[-\gamma_E-\log(m \epsilon)]~. \label{eq:gfour}
\eeq
Remembering that the UV cutoff scale is $\Lambda_{UV}=1/\epsilon$ we see that the approximate CFT corresponding  to this dilaton background contributes to the running of the gauge coupling from the cutoff $\Lambda$ down to a scale
$m\, e^{\gamma_E}$.  This is just what we expect if the CFT unparticles that couple to the external gauge boson have a threshold near $\mu=m\, e^{\gamma_E}$.  In other words the space is effectively cutoff near $z=1/\mu=e^{-\gamma_E}/m$ and
the 4D gauge coupling is
\beq
\frac{1}{g_4^2}\approx \frac{R}{g_5^2}\log\left(\frac{\Lambda_{UV}}{\mu}\right).
\label{4Dgaugecoupling}
\eeq
If we took $\mu\rightarrow 0$ we would find that the 4D gauge coupling ran to zero and the gauge boson zero mode would be non-normalizable.

The equation of motion for the gauge field in this dilaton background is
\beq
f''(z)- 
\left( m + \frac{1}{z} \right) \,f'(z)+p^2 f(z)  =0~.
\eeq
This equation can be solved in terms of confluent hypergeometric functions (more details on the solution can be found in Appendix~\ref{app:gauge}).  In the large $z$ region it is easy to see that the approximate solution is exponential for $0<p<m/2$ and oscillatory for $p>m/2$.
Therefore we expect the two point function of the corresponding CFT operator to have a massless pole and a cut for $p>m/2$  corresponding to the logarithmic running of the gauge coupling for large momenta where the conformal invariance is recovered.
In order to see this, we can study the spectrum of the gauge field in presence of an IR brane at $z=z_{IR}\gg 1/m$.
The spacing between the zero mode and the first KK mode is controlled by the value of $m$; the first KK mode appears near $p=m/2$ while the spacing of the remaining higher KK modes is determined by the infrared cutoff
$z=z_{IR}$.  The low-lying KK modes should correspond to the spin one bound states of the massive degrees of freedom. The spacing of subsequent modes is, as usual, 
$\Delta p\sim 1/z_{IR}$.  Since the 4D gauge coupling (\ref{4Dgaugecoupling}) is independent of $z_{IR}$ if $z_{IR}\gg 1/m$ it is clear that we can take $z_{IR}$ to be arbitrarily large and recover the continuum above $m/2$.

We can now consider the interactions of the gauge zero mode with the fields in the holographic boundary action.
4D gauge invariance requires that we replace the 4D derivative in the EOMs with
\beq
\partial_\mu \to \partial_\mu - i g_4 T^a A_\mu~,
\eeq
where $g_4$ is the 4D coupling defined in Eq.~\eq{gfour}.
This corresponds to minimal gauging of the holographic Lagrangians, and allows one to calculate vertices with an arbitrary number of gauge bosons. 
To calculate the effective vertices, one should solve the bulk EOMs of the matter field including the bulk interactions with the gauge zero mode, which can be considered as a background field because its wave-function is constant in the extra dimension.
To find the single  gauge boson vertex, however, we can use the solution for the free bulk EOM: since the bulk gauge interaction  is already first order term in the gauge field. So we can simply propagate these boundary fields into the bulk and integrate over bulk gauge interaction.  
This integral will give the exact gauge vertex for the holographic action.
More explicitly, in the case of a bulk scalar, we must calculate the integral
\beq
\int dz\frac{d^4p}{(2\pi)^4} \frac{d^4q}{(2\pi)^4}  \left(\frac{R}{z}\right)^{3}\!\!\left(  \partial_M 
\phi^\dagger(p+q,z) T^a A^{a M}(q,z)   \phi(p,z)+h.c.\right)~.
\eeq
This gives the interaction term with the gauge zero mode to be:
\beq
S_{\rm int}=
\int \frac{d^4p}{(2\pi)^4} \frac{d^4q}{(2\pi)^4}   A^{a \alpha}(q)  \int_\epsilon^\infty dz\left(\frac{R}{z}\right)^{3}\left(  \partial_\alpha \phi^\dagger(p+q,z) T^a  \phi(p,z)+h.c.\right)
\eeq
which just yields the usual unparticle vertex \cite{ColoredUnparticles} for the canonically normalized gauge field after we expand for small momenta ($p \epsilon, q \epsilon \ll 1$):
\beq
i g \Gamma^{a\alpha}(p,q)&\equiv&\frac{i\delta^3 S_{\rm int}}{\delta A^{a\mu}(q)\delta \phi_0^\dagger(p+q)\delta \phi_0(p)} \nonumber\\
&=&
i g_4 T^a \frac{2 \sin d_s \pi}{A_{d_s}}  \frac{(2p+q)^\alpha}{2p \cdot q+q^2}\left[ \left(\mu^2-(p+q)^2\right)^{2-d_s}
-\left(\mu^2-p^2\right)^{2-d_s} \right]~.
\eeq

\section{Unparticle Production Cross-sections}
\setcounter{equation}{0}

In this section we want to discuss the unparticle production cross-section via gauge interactions: this would be especially relevant for the LHC in the case of colored unparticles~\cite{ColoredUnparticles}.
A simple way to do the calculation is to gauge the holographic boundary action and calculate the imaginary part of unparticle loop contributions to the $qq \to qq$ scattering: this would give the inclusive cross-section of $q q \to$ unparticles (a calculation along similar lines can give the gluon fusion contribution).
If we neglect the contribution of the logarithmic running of the gauge boson coupling (the IR cutoff $m$ can be arbitrary large), the result for scalar unparticles of dimension $d_s$ is simply given by \cite{ColoredUnparticles}
\beq
\sigma_{d_s} = (2-d_s) \sigma_1\,, \label{eq:scalarXsec}
\eeq
where $\sigma_1$ is the ordinary cross-section into a pair of scalar particles (limit $d_s\to 1$).
This result makes sense only for $d_s<2$: if we naively extended it to larger dimensions we would find a negative cross-section.

This paradox is strictly related to the fact that if we wish to extend the unparticle analysis to $d_s\ge 2$ we find that the spectral density integration in Eq. \eq{unparticle} is divergent. 
In order to get the finite result in Eq. \eq{unparticle} with the proper scaling, we need to subtract some terms from the integral. 
For instance, for $2<d_s<3$ we must subtract a constant term:
\beq
\Delta(p,2< d_s <3)&=&  \frac{A_{d_s}}{2\pi}  \int_{0}^\infty (M^2)^{d_s-2}\left( \frac{i }{p^2-M^2+i \epsilon}- \frac{i }{-M^2}\right)  dM^2 \nonumber\\
&=& \frac{A_{d_s}}{2\pi}  (p^2+i \epsilon)\int_{0}^\infty (M^2)^{d_s-3}\frac{i }{p^2-M^2+i \epsilon} dM^2\nonumber \\
&=&   \frac{A_{d_s}}{2 \sin (d_s-1) \pi} (p^2+i \epsilon) \frac{i}{\left( -p^2-i \epsilon \right)^{2-(d_s-1)} }\nonumber \\
&=&   \frac{A_{d_s}}{2 \sin d_s \pi} \frac{i}{\left( -p^2-i \epsilon \right)^{2-d_s} }\,.
\label{unparticle23}
\eeq
In general, for larger dimensions we need to subtract a  series of terms with increasing powers of $p^2$:
\beq
\frac{A_{d_s}}{2\pi}  \int_{0}^\infty (M^2)^{d_s-2}\left( \frac{i }{p^2-M^2+i \epsilon}- \frac{i }{-M^2}-\frac{-ip^2 }{(-M^2)^2} -\ldots \right)  dM^2\,.
\label{unparticle34}
\eeq
Thus, in order to get the unparticle propagator with the proper scaling for $d_s>2$, we need to subtract a local (divergent) piece from the integral in Eq. \eq{unparticle}:
\beq
\Delta_{\rm local}(p) &=& \frac{A_{d_s}}{2\pi}  \int_0^{\Lambda^2} (M^2)^{d_s-2}\left( \frac{i }{-M^2}+\frac{-ip^2 }{(-M^2)^2} +\ldots \right)  dM^2 \nonumber \\
&=& \frac{i A_{d_s}}{2\pi}\, \Lambda^{2d_s-4} \, \left(    \frac{1}{d_s-2}- \frac{ p^2 \Lambda^{-2} }{d_s-3} +\ldots \right)\,,  
\eeq
where we have imposed a UV cutoff on the momentum integration.

This corresponds directly with what happens in the 5D picture: the scalar 2 point function can be split into a local and non-local term
\beq
\Delta_s (p) = \Delta_{\rm local} (p) + \Delta_{\rm non-local} (p)\,, \label{eq:locnonloc2}
\eeq
in analogy with Eq.~\eq{locnonloc} for fermions.
For $d_s>2$ the local terms are divergent as the UV brane is removed, $\epsilon \sim 1/\Lambda \to 0$.
Thus  for $1<d_s<2$ we can write an effective action for scalar unparticles, which in insensitive to cutoff effects, using the propagator \eq{unparticle}. The fact that conformal symmetry is broken in the UV is unimportant, and all the local terms associated with such breaking can be removed by sending the UV cutoff to infinity. For $d_s>2$ however, it is not possible to consider the conformal sector alone, without including the local terms in the propagator, which are sensitive to cutoff behavior.
In this case the theory consists of a  continuum of states (unparticles), peaked around a mass scale $1/\epsilon$, plus some local contact terms induced by the local part of the propagator in Eq.\eq{locnonloc2}.
If one removes the UV cutoff, i.e. $\epsilon \to 0$, the ($d_s>2$) unparticles decouple: this explains why for $d_s \to 2$ the cross-section goes to zero (or to a small number suppressed by the UV cutoff).
For $d_s>2$, we expect the cross-section to remain small and suppressed by powers of $\epsilon$.
In order to see this, we need to consider the full propagator in Eq. \eq{locnonloc2} (while in Ref.~\cite{ColoredUnparticles} the non-local term only was considered).
The interaction vertex with one gauge boson would be 
\beq
\Gamma^\mu (p,q) &\propto& \left( \Delta_s^{-1} (p+q) - \Delta_s^{-1} (p) \right) 
\simeq  \\
&&  \left( \Delta_{{\rm local}}^{-1} (p+q) - \Delta_{{\rm local}}^{-1} (p) \right) 
- \left(\frac{\Delta_{{\rm non-local}}(p+q)}{ \Delta^2_{{\rm local}} (p+q)} - \frac{\Delta_{{\rm non-local}}(p)}{ \Delta^2_{{\rm local}} (p)} \right) + \dots\,, \nonumber
\eeq
where in the expansion we have used the fact the the local terms are dominant.
If we want to calculate the cross-section for the process $q \bar q \to$ unparticles we need to calculate the imaginary part of the gauge boson propagator dressed with a loop of unparticles. Such imaginary parts can come from taking either local or non-local terms in the propagator and the vertex: however, at leading order when taking the local terms only, there is no imaginary part because the propagators do not have poles and the vertices are real.
Thus, the dominant term will come from taking the local part of the propagator and the non local part of the vertex or vice-versa. 
The dominant contribution to the imaginary part comes from terms like
\beq
\left( \frac{\Delta_{{\rm non-local}}}{\Delta_{{\rm local}}} \right)^2 \sim (\epsilon^2)^{2d_s-4}\,.
\eeq 
This is telling us that, for $d_s$ close to 2 and $d_s>2$, Eq. \eq{scalarXsec} should be replaced by a more complicated function of the cutoff, which however is very suppressed if the cutoff is much larger  than the energy of the experiment. Thus, in first approximation, it is correct to take Eq.~\eq{scalarXsec} for $d_s<2$ and zero for $d_s>2$. Trying to extrapolate Eq.~\eq{scalarXsec} to $d_s>2$ is, as we have shown, meaningless\footnote{This contradicts what is advocated in Ref. \cite{Liao:2007fv}.}.

\subsection{Fermion unparticle cross-section}

\begin{figure}[t]
\centering
\includegraphics[width=0.7\textwidth]{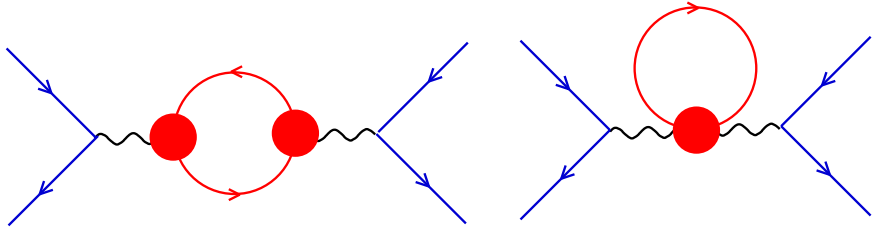}
\caption{Forward scattering of $q{\bar q}$ through a gluon with a fermion unparticle vacuum polarization
loop.  The imaginary parts of those 
diagrams contribute to the inclusive unfermion production cross-section. }
\label{fig:diagrams}
\end{figure}

The calculation of the inclusive $q \bar q \to$ fermion unparticle cross-section is very similar to the scalar case in~\cite{ColoredUnparticles}, the only difference being that the  vertices are more complicated.
The same two types of diagrams in Fig.~\ref{fig:diagrams} contribute due to the fact that the non-local action contains couplings with an arbitrary number of gauge bosons.

The fermion effective action can be written as (see Eq.~\eq{holof}):
\beq
S = \frac{2 \cos d_f \pi}{A_{d_f-1/2}}\, \int \frac{d^4 p}{(2\pi)^4}\;  \bar{\psi}  ( \not p - \mu)\;   [\mu^2 - p^2]^{3/2-d_f} \psi~. \label{eq:actionferm}
\eeq
After gauging such non-local action, we find the following vertex with one gauge boson
\beq
i g\, \Gamma_f^{a\nu}(p,q)
&=&
ig\, T^a\, \gamma^\nu \; \frac{2 \cos d_f \pi}{A_{d_f-1/2}} \frac{1}{2} \left[ \left(\mu^2 - (p+q)^2 \right)^{3/2-d_f} + \left(\mu^2 - (p)^2\right)^{3/2-d_f} \right] + \nonumber \\
&& ig\, T^a\;  \left(\not p +\frac{\not q}{2} -\mu\right)\, (2p^\nu+q^\nu)\, \mathcal{F}_f (p, q)~,
\label{eq:vertex}
\eeq
where
\beq
\mathcal{F}_f (p, q) =\frac{2 \cos d_f \pi}{A_{d_f-1/2}} \frac{\left(\mu^2- (p+q)^2\right)^{3/2-d_f} -\left(\mu^2- (p)^2\right)^{3/2-d_f} }{2 p\cdot q + q^2}~.
\eeq
The vertex with two gauge bosons is more complicated, and we relegated it to the Appendix~\ref{app:Feynman}, where a more general discussion of unparticle gauge vertices can also be found.

It is easy to check that such vertices do respect the (generalized) Ward-Takahashi identities \cite{wt} and they reduce to the usual vertices in the $d_f \to 3/2$ limit (in particular the two gauge boson vertex vanishes - see Appendix~\ref{app:Feynman}).

In the massless limit, $\mu \to 0$, the imaginary part of the two diagrams give the following contributions to the cross-section:
\beq
\left. \frac{\sigma_{d_f}}{\sigma_{F}} \right|_{\rm diag 1} & = & \frac{1}{3} d_f^4 - \frac{8}{3} d_f^3 + \frac{31}{6} d_f^2 - \frac{5}{6} d_f - \frac{33}{16}~,   \\
\left. \frac{\sigma_{d_f}}{\sigma_{F}} \right|_{\rm diag 2} & = & -\frac{1}{3} d_f^4 + \frac{8}{3} d_f^3 - \frac{20}{3} d_f^2 + \frac{19}{3} d_f - \frac{29}{16}~,
\eeq
where we have normalized the results with the cross-section into two massless fermions, $\sigma_{F}$.
In the limit $d_f \to 3/2$, the first diagram will reproduce the usual result, while the second one vanishes.
Summing the two contributions we obtain
\beq
\sigma_{d_f} (q \bar q \to \mbox{unfermions}) = \left( - \frac{3}{2} d_f^2 + \frac{11}{2} d_f - \frac{31}{8} \right) \times \sigma_{F} (q \bar q \to f \bar f)\,. \label{eq:fermXsect}
\eeq
As in the scalar case, a non-trivial cancellation between the two diagrams removes higher powers in $d_f$.
However, unlike in the scalar case, the cross-section does not vanish when the dimension reaches the critical value $d_f \to 5/2$, but it is actually reduced to a half of the particle one.
In the scalar case the cross-section vanishes for $d_s = 2$ because the scalar unparticle becomes 
non-dynamical and all the couplings with gauge bosons vanish individually: this is not the case for fermions due to the non trivial spinorial structure of the couplings and of the kinetic term.
However, when $d_f$ approaches the value $5/2$, the propagator becomes more and more UV sensitive and the action in Eq.~\eq{actionferm} cannot be used anymore: the local terms in the propagator suppress the cross-section more than Eq.~\eq{fermXsect}, as we explained in the scalar case.

\section{Conclusions}
\setcounter{equation}{0}

Unparticles were originally proposed as a candidate for new physics at the LHC that couples to the SM  only through irrelevant operators.
The possibility that unparticles carry SM gauge quantum numbers is particularly interesting, as it can enhance the production cross-section via strong or weak interactions (compared to higher dimension operators) and give rise to new interesting signals in the ATLAS and CMS detectors (compared to just missing energy).
However, the early work on unparticle physics left many theoretical issues that had not been sufficiently clarified in the literature.
For large scaling dimensions ($d_s > 2$ for scalars and $d_f > 5/2$ for fermions) the unparticle  propagator is UV sensitive.
Even though the resulting phase space is regular, other calculations that make use of such a propagator can give nonsensical results unless care is taken.

Comparison between the  AdS description of a CFT and the unparticle description allows for a simple way to  deal with complicated issues like gauge interactions for unparticles. 
In this paper we showed that AdS theories without an IR cutoff are equivalent to unparticles, in the sense that CFT operators corresponding to bulk fields have the same scaling properties and propagators as unparticles. It could not have been otherwise.
Moreover, we showed that holographic boundary actions can be used as effective actions for unparticles, and that they reproduce the proposal in \cite{ColoredUnparticles} for scalars with $1 < d_s < 2$ (and, in the case of fermions, $3/2 < d_f < 5/2$).
For larger scaling dimensions, local terms in the boundary action  appear that are enhanced by powers of the UV cutoff $\Lambda$ (which is inversely proportional to the position $\epsilon$ of the UV brane in the extra dimension).
Such terms correspond to the UV sensitivity of unparticle propagators.  These terms do not contribute to the final state phase space, however they cannot be neglected in the propagator and they give rise to effective contact interactions.
We also showed that, taking them into account, the calculation of the production cross-section via gauge interactions can be extended to $d_s > 2$ and that the cross-section is cutoff suppressed in this regime.
We also presented the result for fermion unparticle production, which shows a very similar behavior.

Unparticles can also shed light on AdS theories.
We used unparticles as a guide to develop new results in AdS field theories.  There is a very simple way to introduce an IR cutoff for unparticles which preserves continuum behavior above a threshold.  We showed how to reproduce this behavior in AdS  with a bulk VEV which grows in the IR. This provides an alternative to the usual Randall-Sundrum constructions with an IR brane producing a hard IR cutoff.  The new IR cutoff  introduces a mass gap, but without the discrete KK tower which is associated with confining theories.
We studied two simple cases of a bulk VEV with a profile growing towards the IR, one with a quadratic profile which produces a threshold for a scalar field and a linear profile for a fermion threshold.  The back-reaction of such a growing VEV will eventually modify the geometry and effectively cut off the space, however this can happen at much lower scales so that  the continuum approximation is still valid.

Finally we addressed the issue of gauge interactions for unparticles. In the AdS description without an IR cutoff, there is no normalizable zero mode, corresponding to the fact that the gauge coupling for a gauge boson which gauges a global symmetry of a large $N$ CFT runs to zero in the IR.
This effect can be thought of as the result of a mixing with CFT states with the same quantum numbers as the gauge boson and with arbitrarily small masses.
However, if we consider an AdS theory with a dilaton factor multiplying the bulk gauge action then if the dilaton profile falls sufficiently fast in the IR the running is effectively cut off at some threshold scale and only the ordinary massless gauge boson is left at low energies.
Moreover we showed that  the couplings of the massless gauge boson to the unparticles (boundary values of bulk fields) will respect the minimal coupling prescription used in \cite{ColoredUnparticles}: we therefore calculated the inclusive fermion unparticle production cross-section (relevant for the LHC) which shows similar properties as the scalar one.

\vspace{24pt}

{\bf Note Added:}  While finishing the write-up of this paper we noticed the paper in Ref.~\cite{otherUnfermions} which attempts to consider gauge interactions of unfermions, however the propagators used in that work are not consistent with those found in the AdS/CFT correspondence. Moreover, the effective action the authors use is different and, in particular, it does not have a $d$-dependent power in the momentum.

\section*{Acknowledgments}
We thank Hsin-Chia Cheng, Markus Luty, Michele Redi, Matt Reece, and Jesse Thaler for useful discussions and comments.
GC and JT thank the Kavli Institute for Theoretical Physics in Santa Barbara, CA for hospitality during completion of this work.
This work is supported in part  by the US department of Energy under contract DE-FG02-91ER406746
and in part by the National Science Foundation under
Grant No. PHY05-51164. 
JT  acknowledges the hospitality and support of the
  Radcliffe Institute for Advanced Study during the completion of this work.

\appendix

\section{Appendix: IR cutoff for fermion fields in AdS}
\label{app:calculation}
\setcounter{equation}{0}

The bulk action for the two bulk fermions is:
\beq 
S_{bulk}  &=& \int d^4 x dz 
\left(\frac{R}{z}\right)^4 
\left\{ \vphantom{\frac{v (z)}{z} }
- i \bar \chi_{L/R}  \bar{\sigma}^\mu \partial_\mu  \chi_{L/R} 
- i \psi_{L/R}  \sigma^{\mu} \partial_\mu \bar{\psi}_{L/R}\right. \nonumber \\ 
&&\quad\quad\quad\quad\quad\quad\quad\,\,
\left. + \sfrac{1}{2} ( \psi_{L/R} \overleftrightarrow{\partial_z}  \chi_{L/R} 
-  \bar{ \chi}_{L/R}  \overleftrightarrow{\pd_z} \bar{\psi}_{L/R} )
+ \frac{c_{L/R}}{z} \left( \psi_{L/R}  \chi_{L/R} + \bar{ \chi}_{L/R} \bar{\psi}_{L/R} \right) \right. \nonumber \\ 
&&\quad\quad\quad\quad\quad\quad\quad\,\,
+\left. \frac{v (z)}{z} \left( \psi_{L}  \chi_{R} + \psi_{R}  \chi_{L} + \bar{ \chi}_{L} \bar{\psi}_{R} +\bar{ \chi}_{R} \bar{\psi}_{L}  \right) \right\}\,.
\eeq 
The EOMs can be solved by

\begin{equation} 
\chi_{L/R} 
= 
g_{L/R}(z)\, \chi_4
\ \ \mathrm{and} \ \ 
\bar{\psi}_{L/R} 
=
f_{L/R} (z)\, \bar{\psi}_4, 
\end{equation} 
where the wave functions $g_{L/R}$ and $f_{L/R}$ are the solutions of
\beq
p \left( \begin{array}{c} g_L \\ g_R \end{array} \right) + \left[ \partial_z - \frac{1}{z} \left( \begin{array}{cc} c_L + 2 & v \\ v & c_R + 2 \end{array} \right) \right] \left( \begin{array}{c} f_L \\ f_R \end{array} \right)=0\,, \label{eq:eom1} \\
p \left( \begin{array}{c} f_L \\ f_R \end{array} \right) - \left[ \partial_z + \frac{1}{z} \left( \begin{array}{cc} c_L - 2 & v \\ v & c_R - 2 \end{array} \right)\right]  \left( \begin{array}{c} g_L \\ g_R \end{array} \right)=0\,. \label{eq:eom2}
\eeq
We can now determine the $f$'s from Eq.~(\ref{eq:eom2}) and plug it back in Eq.~(\ref{eq:eom1}) to find second order equations for the $g$'s:
\beq
\left[ \partial_z^2 - \frac{4}{z} \partial_z + p^2 - \frac{1}{z^2} \left( \begin{array}{cc} v^2 + c_L^2 - c_L - 6 & v (c_L + c_R) + v - v' z \\  v (c_L + c_R) + v - v' z &  v^2 + c_R^2 - c_R - 6 \end{array} \right) \right]  \left( \begin{array}{c} g_L \\ g_R \end{array} \right) = 0\,.
\eeq
If $c_L = - c_R\equiv c$ and if the VEV is linear in $z$, $v(z) = \mu z$, then the two equations decouple:
\beq
\left[ \partial_z^2 - \frac{4}{z} \partial_z + p^2 - \mu^2 - \frac{c^2 \mp c - 6 }{z^2}\right] g_{L/R} = 0\,.
\eeq
A similar equation can be derived for the $f$'s:
\beq
\left[ \partial_z^2 - \frac{4}{z} \partial_z + p^2 - \mu^2 - \frac{c^2 \pm c - 6 }{z^2}\right] f_{L/R} = 0\,.
\eeq

Defining $E = \sqrt{p^2 - \mu^2}$, the bulk solutions are:
\beq
g_{L,R} & = & z^{5/2} \left( a_{L/R} J_{\pm c + 1/2} (E z) + b_{L/R}  J_{\mp c - 1/2} (E z) \right)\,, \\
f_{L,R} & = & z^{5/2} \left( a'_{L/R} J_{\pm c - 1/2} (E z) + b'_{L/R}  J_{\mp c + 1/2} (E z) \right)\,.
\eeq
Imposing the first order Eqs.~(\ref{eq:eom1}) (or equivalently Eqs.~(\ref{eq:eom2})), we can determine half of the parameters in the solutions:
\beq
\left\{ \begin{array}{l} p b_L + E b'_L - \mu a'_R = 0 \\ p a_L - E a'_L - \mu b'_R = 0 \end{array} \right. \quad& \Rightarrow &\quad \left\{ \begin{array}{l} a_L = \frac{p a'_L - \mu b_R}{E} \\ b_L = - \frac{p b'_L - \mu a_R}{E} \end{array} \right. \\
\left\{ \begin{array}{l} p b_R + E b'_L - \mu a'_L = 0 \\ p a_R - E a'_R - \mu b'_L = 0 \end{array} \right. \quad& \Rightarrow &\quad \left\{ \begin{array}{l} a'_R = \frac{p a_R - \mu b'_L}{E} \\ b'_R = - \frac{p b_R - \mu a'_L}{E} \end{array} \right.
\eeq
The BCs on the UV brane

$$
\psi_L (\epsilon) = \psi_0 = f_{0} \psi_4\,, \quad \mbox{and} \quad \chi_R (\epsilon) = \chi_0 = g_{0} \chi_4\,,
$$
will determine the overall normalizations
\beq
\left( \begin{array}{c} a_R \\ b_R \end{array} \right) &=& \frac{\epsilon^{-5/2} g_0}{ s_{\alpha R} J_{c-1/2} (E \epsilon)+c_{\alpha R} J_{-c+1/2} (E \epsilon)} \left(  \begin{array}{c} c_{\alpha R}\\ s_{\alpha R} \end{array} \right)  \\
\left( \begin{array}{c}a'_L \\ b'_L \end{array} \right) &=& \frac{\epsilon^{-5/2} f_0}{ c_{\alpha L} J_{c-1/2} (E \epsilon)+s_{\alpha L} J_{-c+1/2} (E \epsilon)} \left(  \begin{array}{c} s_{\alpha L}\\ c_{\alpha L} \end{array} \right)
\eeq
with $c_{\alpha L}^2 + s_{\alpha L}^2 =1 = c_{\alpha R}^2 + s_{\alpha R}^2 $.

After imposing the EOMs in the bulk, we are left with a boundary action
\beq
S_{\rm boundary} = \int d^4 x \left( \frac{R}{\epsilon} \right)^4 \frac{1}{2} \left( g_L (\epsilon) f_L (\epsilon) + g_R (\epsilon) f_R (\epsilon) \right) \left[ \psi_4 \chi_4 + \bar{\chi}_4 \bar{\psi}_4 \right]\,.
\eeq
Plugging in the solutions, we obtain Eq.~(\ref{eq:LbraneIR}).

\section{Appendix: IR threshold in the gauge sector}
\label{app:gauge}
\setcounter{equation}{0}

In presence of the dilaton factor in Eq.~\eq{dilaton}, the EOM for the gauge boson becomes
\beq
f''(z)- 
\left( \frac{1}{z}- \frac{\Phi'}{\Phi} \right) \,f'(z)+p^2 f(z)  =0~.
\eeq
If the dilaton decreases with a power of $z$, the solutions will still be Bessel functions and generate a continuum all the way down to zero momentum.
Therefore we need a faster decrease, an exponential, in order to get a completely different spectrum.
If $\Phi = e^{-m z}$, the EOM is
\beq
f''(z)- 
\left( \frac{1}{z} + m \right) \,f'(z)+p^2 f(z)  =0~;
\eeq
for $z \ll 1/m$ the solutions are exponential functions, while for $z \gg 1/m $ we obtain the usual combination of Bessel functions.
The equation can be solved analytically: first we can rescale the solution
\beq
f (z) = z^2 e^{\frac{z}{2} \left( m - \sqrt{m^2 - 4 p^2} \right)} g(z)~, \label{eq:prefactor}
\eeq
such that $g$ satisfies the following equation:
\beq
\frac{z}{\sqrt{m^2-4 p^2}} g''(z) + \left( \frac{3}{\sqrt{m^2-4 p^2}} - z \right) g'(z) - \left( \frac{3}{2} + \frac{m}{2 \sqrt{m^2 - 4 p^2}} \right) g(z) = 0~.
\eeq
After a change of variables $y =\sqrt{m^2 - 4 p^2}\; z$, we recognize this as a confluent hypergeometric (or Kummer) differential equation:
\beq
y g'' + (3-y) g' - \alpha g =0~, \quad \mbox{where} \quad \alpha =  \frac{3}{2} + \frac{m}{2 \sqrt{m^2 - 4 p^2}}~,
\eeq
whose solution is a combination of confluent hypergeometric functions.

The form of the spectrum can be inferred by the exponential prefactor in Eq.~\eq{prefactor}: for $p<m/2$ it is a real exponential, therefore only the zero mode is present at low energies; for $p>m/2$, the square root develops an imaginary part and the solution becomes oscillatory, therefore producing, above $m/2$, a tower of KK states with spacing given by the IR brane.

\section{Appendix: summary of Feynman rules for unparticle gauge interactions}
\label{app:Feynman}
\setcounter{equation}{0}

In this Appendix we will summarize the Feynman rules for unparticle gauge interactions.
We will start with the scalar case, already discussed in ~\cite{ColoredUnparticles}.
The effective action for $1 \leq d_s < 2$ is
\beq
S_{\rm scalar} = \frac{2 \sin d_s \pi}{A_{d_s}} \int \frac{d^4 p}{(2 \pi)^4}\; \phi^\dagger (-p) \left[ \mu^2 - p^2 \right]^{2-d_s} \phi (p)~.
\eeq
Minimal gauging applied to this non-local action~\cite{gauging,gauging2} gives the following vertex with one gauge boson:
\beq
i g\, \Gamma_s^{a\nu} (p, q) = i g\, T^a  (2p^\nu+q^\nu)\, \mathcal{F}_s (p, q)~,
\eeq
where the dependence on the unparticle propagator is embedded in the function $\mathcal{F}$, defined for scalars as:
\beq
\mathcal{F}_s (p,q) = \frac{2 \sin d_s\pi}{A_{d_s}}\; \frac{\left( \mu^2 - (p+q)^2 \right)^{2-d_s} -\left( \mu^2 - p^2 \right)^{2-d_s}}{2 p\cdot q + q^2}~.
\eeq
The vertex with two gauge bosons is $i g^2\, \Gamma_s^{a\mu,b\nu}(p,q_1,q_2)$, with
\beq
\Gamma_s^{a\mu,b\nu}
&=& (T^a T^b + T^b T^a)\,  g^{\mu \nu} \mathcal{F}_f (p, q_1+q_2) + \nonumber \\
& & T^a T^b\, \frac{(2 p^\nu + q_2^\nu) (2 p^\mu + q_1^\mu + 2 q_2^\mu)}{q_1^2 + 2 p\cdot q_1 + 2 q_1\cdot q_2}\; \left( \mathcal{F}_f (p, q_1+q_2) - \mathcal{F}_f (p, q_2) \right) + \nonumber \\
& &  T^b T^a\, \frac{(2 p^\mu + q_1^\mu) (2 p^\nu + q_2^\nu + 2 q_1^\nu)}{q_2^2 + 2 p\cdot q_2 + 2 q_1\cdot q_2}\; \left( \mathcal{F}_f (p, q_1+q_2) - \mathcal{F}_f (p, q_1) \right)~.
\eeq
In the $d_s \to 1$ limit, $\mathcal{F}_s \to 1$, and we recover the usual scalar vertices
\beq
i g\, \Gamma_{s(d_s=1)}^{a\mu} = i g\, T^a (2p^\mu + q^\mu)~, \quad \mbox{and} \quad i g^2\, \Gamma_{s(d_s=1)}^{a\mu,b\nu} = i g^2\, (T^a T^b + T^b T^a) g^{\mu \nu}~.
\eeq

In the fermion case, the effective action for $3/2 \leq d_f < 5/2$ is
\beq
S_{\rm fermion} = \frac{2 \cos d_f \pi}{A_{d_f-1/2}}\, \int \frac{d^4 p}{(2\pi)^4}\;  \bar{\psi} (-p)\,  ( \not p - m)   [m^2 - p^2]^{3/2-d_f}\, \psi (p)~.
\eeq
The gauge vertices are more complicated, however they can be written in terms of the scalar ones.
The correct dependence on the fermion dimension is recovered replacing $\mathcal{F}$ with the following function
\beq
\mathcal{F}_f (p, q) =\frac{2 \cos d_f \pi}{A_{d_f-1/2}}\; \frac{\left(\mu^2- (p+q)^2\right)^{3/2-d_f} -\left(\mu^2- (p)^2\right)^{3/2-d_f} }{2 p\cdot q + q^2}~.
\eeq
For the one gauge boson vertex the result is
\beq
i g\, \Gamma_f^{a\nu}(p,q)
&=&
ig\, T^a \gamma^\nu \, \frac{2 \cos d_f \pi}{A_{d_f-1/2}} \frac{1}{2} \left[ \left(\mu^2 - (p+q)^2 \right)^{3/2-d_f} + \left(\mu^2 -p^2\right)^{3/2-d_f} \right]  \nonumber \\
&&  +\left(\not p +\frac{\not q}{2} -\mu\right)\; i g\, \Gamma_s^{a\nu} (p,q)~.
\eeq
The two gauge boson vertex can also be written in terms of scalar vertices
\beq
i g^2\, \Gamma_f^{a\rho,b\nu}(p,q_1,q_2)
&=&  \left( \not p + \frac{{\not q}_1 + {\not q}_2}{2} -\mu \right)\;i g^2\, \Gamma_s^{a\rho,b\nu} (p, q_1, q_2)  \nonumber\\ 
& &+ \frac{1}{2} \gamma^\rho\;i g^2\, \Gamma^{b\nu,a} (p, q_2, q_1) +  \frac{1}{2} \gamma^\nu\; i g^2\, \Gamma^{a\rho,b} (p, q_1, q_2)~,
\eeq
where  $\Gamma^{a\nu,b}$ is given by:
\beq
\Gamma^{a\nu,b}(p,k_1,k_2)
&=& T^a T^b \left( 2 p^\nu + k_1^\nu\right) \mathcal{F}_f (p, k_1) + \nonumber \\
& & T^b T^a \left( 2 (p^\nu + k_2^\nu) + k_1^\nu \right) \mathcal{F}_f (p+k_2, k_1)~.
\eeq
In the particle limit, $d_f \to 3/2$, we find that $\mathcal{F}_f \to 0$ and we easily recover the usual fermion couplings
\beq
i g\, \Gamma_{f(d_f=3/2)}^{a\nu} = i g\, T^a\, \gamma^\nu~, \quad \mbox{and} \quad i g^2\, \Gamma_{f(d_f=3/2)}^{a\mu,b\nu} = 0~.
\eeq 

As a non-trivial check of our results, we can verify that the vertices do respect the Ward-Takahashi identities \cite{wt} in the abelian case:
\beq
i g\, q_\mu \Gamma_{s/f}^\mu (p,q)  & = & g \left( \Delta^{-1}_{s/f} (p) - \Delta^{-1}_{s/f} (p+q) \right)~, \\
i g^2\, q_{1\mu} \Gamma_{s/f}^{\mu \nu} (p,q_1,q_2) &=& g \left( i g\, \Gamma_{s/f}^{\nu} (p+q_1, q_2) - i g\, \Gamma_{s/f}^{\nu} (p, q_2) \right)~,
\eeq
where we have suppressed the group indices and set $T^a = 1$.
In the non-abelian case, the vertices respect more complicated relations:
\beq
i g\, q_\mu \Gamma_{s/f}^{a\mu} (p,q)  & = & g\, T^a \left( \Delta^{-1}_{s/f} (p) - \Delta^{-1}_{s/f} (p+q) \right)~, \\
i g^2\, q_{1\mu} \Gamma_{s/f}^{a\mu, b\nu} (p,q_1,q_2) &=& g \left( i g\, \Gamma_{s/f}^{b\nu} (p+q_1, q_2)\, T^a - i g\, T^a\, \Gamma_{s/f}^{b\nu} (p, q_2) \right) + \nonumber \\
& & g f^{abc}\; i g\, \Gamma_{s/f}^{c\nu} (p, q_1+q_2)~.
\eeq

\end{document}